\documentstyle[epic,eepic,12pt]{article}
\begin{document}
\bibliographystyle{unsrt}
\newcommand{\boldr}{\mbox{${\bf r}$}}
\title{ The Aqueous Solvation of Water: A Comparison of Continuum Methods with
Molecular Dynamics}
\author{ Steven W. Rick and B. J. Berne\\
Department of Chemistry and Center for Biomolecular Simulation\\
Columbia University, New York, NY 10027 }
\maketitle
\begin{abstract}
The calculation of the solvation properties of a single water molecule
in liquid water is carried out in two ways.
In the first, the water molecule is placed in a cavity and the solvent
is treated as a dielectric continuum. This model is analyzed by numerically
solving the Poisson equation using the DelPhi program.
The resulting solvation properties depend sensitively on the shape
and size of the cavity. In the second method, the solvent and solute
molecules are treated explicitly in molecular dynamics simulations
using Ewald boundary conditions. We find a 2 kcal/mole difference in solvation
free energies predicted by these two methods when standard cavity
radii are used.
In addition,
dielectric continuum theory assumes that the solvent reacts solely
by realigning its electric moments linearly with the strength of the solute's
electric field;
the results of the molecular simulation show important
non-linear effects.
Non-linear solvent effects are generally of two types: dielectric
saturation, due to solvent-solute hydrogen bonds, and electrostriction,
a decrease in the solute cavity due to an increased electrostatic
interaction.
We find very good agreement between the two methods if the
radii defining the solute cavity used in the continuum theory
is decreased with the solute charges,
indicating that electrostriction is the
primary non-linear effect and suggesting a procedure for improvement of
continuum methods.
The two methods cannot be made to agree when the atomic radii are made
charge independent, but charge dependent cavity radii are shown to greatly
improve agreement.
\end{abstract}
\noindent
\baselineskip=1.8\baselineskip
\eject
{\bf Introduction}

Dielectric continuum methods based on the numerical solution of
the Poisson Equation provide estimates of solvation free energies.
Because this approach is much faster than
full molecular dynamics simulations, continuum theory (CT)
has been applied to a wide variety of
systems, from aqueous solutions of
large solute molecules such as proteins and nucleic acids to
small solutes such as atomic ions \cite{SharpHonig}.
The widespread use of such continuum calculations makes
it important to test them against simulations
with explicit solvent molecules
and against experimental data where available.
Previous comparisons of continuum theory
and molecular simulations have found good
agreement for solvation energies, typically about 10{\%}
\cite{Jean-Charles,Jayaram}.
The focus of this note will be to compare continuum theory with
full molecular simulations of a small polar solute, for
equilibrium properties, such as
solvation energies, electrostatic potentials,
and average electric fields, from which the
electrostatic forces can be calculated.

The solvation process studied here is the solvation of a single water molecule
in liquid water.
The free energy of solvation, $\Delta {\rm A}_{sol}$, of a water molecule can
be decomposed into a three-step process: 1) removal of the gas-phase partial
charges from the gas-phase water molecule ($\Delta {\rm A}_{es}^{gas}$);
2) insertion of
the uncharged water molecule into the solvent making a hydrophobic cavity
in the bulk water ($\Delta {\rm A}_{cav}$); and
3) charging the water solute to the desired liquid phase
charges ($\Delta {\rm A}_{es}$) (see Figure \ref{fig:cycle}) \cite{Sharp}.
The free energy, $\Delta {\rm A}_{es}^{gas}$, for the first step is the
self-energy for electronic polarization in the gas-phase or the free
energy difference between the isolated molecule with its gas phase partial
charges and with no partial charges.
The free energy for the liquid-state charging step,
$\Delta {\rm A}_{es}$, contains two
parts, a self-energy for polarization in the liquid phase
and a solvent electrostatic contribution.
The two polarization terms will not completely cancel
since the gas phase and solution
charges are not necessarily (or generally) equal and also the
polarization energy will depend on the medium.
The two polarization energies are commonly neglected
[but see \cite{Sharp,Berendsen}].

The solvent electrostatic contribution to the free energy is the
focus of this paper and is the subject of comparison
between the full molecular simulation and the
continuum calculations.
This step consists of reversibly charging
the hydrogen charge, Q$_H$, on the solute water molecule
from 0 to 0.5e
(the oxygen charge, Q$_O$, equals -2Q$_H$) while
keeping the solvent charges equal to the usual liquid state values.
(We are including charges higher than Q$_H^{liq}$---higher than is
necessary for the evaluation of $\Delta {\rm A}_{es}$ for
${\rm H}_2 {\rm O}$---in order to look at the effects of high charge.)
In the continuum approximation, all the solvent electrostatic
properties are contained in the dielectric constant, $\epsilon$.
Setting $\epsilon$ equal to the same value given by the molecular
potential used in the simulation, we can assure self-consistency.

The properties of interest are the energy of the solute, the average
electrostatic potential and the field of the solvent
 at the solute charge sites.
The electrostatic potential, $\phi$, at
position $\boldr_{1\alpha}$ (where 1 labels the solute
molecule and $\alpha$ labels the charge site) due to the solvent is given by
\begin{equation}
\phi({\bf r}_{1 \alpha}) = \sum_{j=2}^{N} \sum_{\beta=1}^3 {\rm Q}_{\beta} /
\left| {\bf r}_{1\alpha} - {\bf r}_{j\beta} \right| .
\label {eq:phi}
\end{equation}
The electrostatic potential energy of the solute is
\begin{equation}
\langle {\rm U}_{es} \rangle = \sum_{\alpha=1}^3 {\rm Q}_{\alpha} \langle
\phi_{\alpha}
\rangle
\label {eq:ues}
\end{equation}
and the electric field is
\begin{equation}
{\bf E}(\boldr_{1\alpha}) = - \left \langle { \partial \phi(\boldr_{1\alpha})
\over
\partial \boldr_{1\alpha} } \right \rangle
\label {eq:field}
\end{equation}
where the brackets $\langle \cdots \rangle$ indicate an average over the
solvent configurations.
The free energy, $\Delta {\rm A}_{es}$, for the solute with a set of
partial charges ${\bf Q}$
can be calculated using the charging integral,
\begin{eqnarray}
\Delta {\rm A}_{es} ({\bf Q} ) & = &
 \sum_{\alpha=1}^3 {\rm Q}_\alpha \int_0^1 \langle \phi
(\boldr_{1\alpha}) \rangle_\lambda d\lambda \nonumber \\
 & = & \int_0^1 \langle {\rm U}_{es} \rangle_{\lambda} /\lambda \, d\lambda
\label {eq:dg}
\end{eqnarray}
where the angular brackets, $\langle \cdots \rangle_\lambda$, indicate an
ensemble average with the solute charges equal to $\lambda {\bf Q}$
\cite{mcquarrie}.
Eq.(\ref{eq:dg}) is integrated numerically by performing
simulations at 11 different hydrogen charges, ranging from 0 to .5.

Continuum theories assume that the solvent response to the
solute is linear in the solute charge. Specifically, the potential,
$\phi(\boldr_{1\alpha})$, is linear in the charge Q$_\alpha$ and therefore,
from Eq.(\ref{eq:dg}), $\Delta {\rm A}_{es} \propto {\rm Q}^2$.
The results of the simulations will test the assumption of linear response.
Specifically, continuum theory based on the Poisson equation is a linear
theory in that it assumes that the
orientations of the solvent molecules respond linearly to
the electric field of the solute \cite{bottcher}.
In real molecular liquids,
non-linear responses not described by the Poisson equation
arise from many factors including
the formation of solute-solvent hydrogen bonds
and electrostriction, the decrease in the excluded volume of the solute
due to the increased solute-solvent Coulombic interaction.
Strong hydrogen bonds prevent the solvent from further
responding to the solute causing dielectric saturation which
decreases the solvation energy.
On the other hand, electrostriction increases the solvation energy.

\vspace{5 mm}
\noindent
{\bf Methods}

The water potential used here is the SPC potential, characterized by
an OH bond length of 1 \AA, an HOH bond angle of 109.47$^\circ$,
charges on the hydrogens and oxygen equal to 0.41e and -0.82e, respectively,
and a Lennard-Jones interaction between oxygen atoms with a well depth,
$\epsilon/k_B$,
equal to 78.2 Kelvin and a radius, $\sigma$, equal to 3.166 \AA
\cite{BerendsenSPC}.
The molecular dynamics simulations were performed on the
Connection Machine CM-5
using a direct ${\rm N}^2$ method\cite{Lynch}, with
511 {\it solvent} molecules and 1
{\it solute} molecule. Periodic boundary conditions, using the Ewald sum
for the long-ranged electrostatic potentials,
a time step of 1 femtosecond and the SHAKE algorithm for enforcing
bond constraints are used{\cite{Allen}}.
The simulations are done in the canonical (constant T,V,N) ensemble
by coupling to a Nos\'{e} thermostat{\cite{Allen,Nose}} and
are at a density of 1 g/cm$^3$ and a temperature
of 300 Kelvin.
Each data point at a given charge represents a 40 picosecond
simulation\cite{footnote1}.
The electrostatic potential with periodic boundary conditions is
\begin{equation}
\phi(\boldr_{1\alpha}) = \sum_{\bf n} {}' \sum_{j=1}^N \sum_{\beta}
{\rm Q}_\beta /
\left| \boldr_{1\alpha} - \boldr_{j\beta} + {\bf n} \right|
\label {eq:phi2}
\end{equation}
where the prime on the sum over periodic images {\bf n} indicates that for
${\bf n}=0$ the term j=1 is omitted. In the standard Ewald evaluation of
Eq. (\ref{eq:phi2}), the energy is written as a sum of four terms,
\begin{eqnarray}
\phi(\boldr_{1\alpha}) &
= & \sum_{j=2}^N \sum_{\beta} {\rm Q}_\beta
{\rm erfc}(\kappa \left| \boldr_{1\alpha} - \boldr_{j\beta} \right| )
/ \left| \boldr_{1\alpha} - \boldr_{j\beta} \right|
\nonumber \\
& +& \sum_{j=1}^N \sum_\beta {\rm Q}_\beta { 4 \pi \over L^3 } \sum_{{\bf G}
\neq 0}
{1 \over G^2 } e^{-G^2/4\kappa^2} \cos({\bf G} \cdot
(\boldr_{1\alpha}- \boldr_{j\beta}))
\nonumber \\
& - &  {2 \pi \over 3 L^3 } \sum_{j=1}^N \sum_\beta {\rm Q}_\beta
(\boldr_{1\alpha}- \boldr_{j\beta})^2
\nonumber \\
& - & \sum_\beta {\rm Q}_\beta
{\rm erf}(\kappa \left| \boldr_{1\alpha} - \boldr_{1\beta} \right| )
/ \left| \boldr_{1\alpha} - \boldr_{1\beta} \right|
\label {eq:Ewald}
\end{eqnarray}
where $\kappa$ is parameter in the Ewald sum chosen for computational
convenience to be 6/L, L is the length of the primary simulation
cell, and {\bf G} is a recriprocal lattice
vector of the periodic simulation cells\cite{Allen}.

In the continuum calculations, the solute is characterized by a
molecular cavity defined by spheres around each charge site and
a dielectric continuum outside this cavity.
The potential, $\phi$, can then be found by solving the
Poisson equation
\begin{equation}
\nabla \cdot \epsilon (\boldr) \nabla
\phi ( \boldr) + 4 \pi \rho ( \boldr) = 0 \label{eq:PB}
\end{equation}
where $\epsilon$(r) is a position dependent
dielectric constant, equal to 1 inside the
solute cavity and 65 outside (65 is the dielectric constant of
the SPC water model\cite{Alper,Belhadi}),
and $\rho(\boldr)$ is the charge density.
Equation (\ref{eq:PB}) is
solved using the DelPhi program, which discretizes space on a
cubic grid (with 65$\times$65$\times$65 points)\cite{Nicholls}.
This approach is thus based on a finite-difference solution of the
Poisson Equation. The calculations reported here use focusing
boundary conditions, in which successive calculations are done, each
with a finer mesh, using the previous coarser grid results to correct
for the long range boundary effects. At the highest resolution, 80$\%$ of the
grid points are inside the solute cavity, corresponding to a
grid spacing of about 0.034 {\AA}.
The water molecule geometry is the SPC geometry.
The oxygen cavity radius, ${\rm r}_O$, is set equal 1.77 {\AA}
and the hydrogen cavity radius is 0.8 {\AA}. The oxygen radius
is about $2^{-5/6}\sigma$, one half the
minimum of the Lennard-Jones potential. This is the standard method
for choosing radii in DelPhi calculations {\cite{Sharp,Jean-Charles}.

\vspace{5 mm}
\noindent
{\bf Results and Discussion}

The free energies corresponding to the molecular dynamics (md)
simulations and to the
CT calculations are
shown in Figure \ref{fig:energy}.
At the charge value of SPC water
(Q$_H$=.41),
the md yields -8.4$\pm$.5
and CT yields -10.5 kcal/mole for $\Delta {\rm A}_{es}$.
The reported error bars are two standard deviations.
This difference of about $\pm$ 2 kcal/mole between the two methods is
comparable to the agreement between the CT and molecular free energy
calculations (using the TIP4P water potential) for several solutes (but not
water)\cite{Jean-Charles}.
The charge dependence for the simulation data is not quadratic, as can be
seen by the poorness of a quadratic fit (the dotted line).

Previous calculations for SPC and also
TIP4P water embedded in a dielectric continuum
find $\Delta {\rm A}_{es}$=-10.96 and -10.89 kcal/mole,
respectively\cite{Rashin}.
These calculations use an oxygen cavity radius
of 1.5{\AA} and a hydrogen cavity radius of 1.16{\AA}.
The sensitivity of the CT results to these input parameters will
be discussed below.
Other CT calculations by Sharp, {\it et al.} \cite{Sharp},
for the TIP4P geometry, which include polarizability of the solute,
find $\Delta {\rm A}_{es}$=-9.3 kcal/mole.
There have been calculations of $\Delta {\rm A}_{sol}$ from molecular
simulations, but the electrostatic part, $\Delta {\rm A}_{es}$, was
not reported so there is no other molecular simulation to compare to
\cite{Mezei}.

Figure \ref{fig:phi} shows
the average electrostatic potentials, $\langle \phi \rangle$.
The md and CT results are qualitatively
different in two respects: 1) $\langle \phi_H \rangle$
and $\langle \phi_O \rangle$ from the md simulations are not
linear in charge
and 2) $\langle \phi_H \rangle$ and $\langle \phi_O \rangle$
from md are not zero at zero charge,
whereas the $\phi$'s from the CT calculations are linear in charge and
zero at zero charge.
As is well known, molecular water solvates a small
uncharged solute by forming a clathrate cage
around the uncharged sphere\cite{Davidson}.
For realistic water potentials, the electrostatic
potential is not zero inside the cage because fluctuations of
the water molecules that form the clathrate cage are known to
bring the hydrogen atoms closer to the cage than the oxygen atoms.
This is apparent from the charge distribution function (see below).
Since the positive charges can get closer than the negative charges there
is a net electrostatic potential inside the sphere.
However, it is the spatial dependence and not the
exact value of $\phi$ that is important because
1) properties such as the electric field and the energy are
invariant with respect to an additive constant in $\phi$
and 2) the exact value of $\langle \phi \rangle$
is strongly dependent on the choice of
boundary conditions used in the Ewald sum.
In the Ewald summation, the periodically replicated system must be
surrounded by a medium, in Eq. (\ref{eq:Ewald}) the medium is taken to be
an insulator (this boundary term is the third term on the
right-hand side of the equation). If the surrounding medium is taken to
a conductor, then this term vanishes\cite{Allen}.
The boundary conditions have only a slight influence on most properties,
however, for the electrostatic potential the boundary conditions give rise to a
large constant term.

The electric field, {\bf E}, arising from the solvent
at each solute site is shown in Figure \ref{fig:field}.
The geometry of the water molecule is shown in the
upper left corner of Figure \ref{fig:field}.
The top panel shows the x-component of the field, ${\rm E}_x$,
at the position of
the hydrogen ${\rm H}_1$, ${\rm E}_x$
at the other hydrogen is minus the
field at ${\rm H}_1$ and ${\rm E}_x$
at the oxygen is zero.
The bottom panel shows the y-component of the field
at the oxygen and hydrogen positions.
Since there is no net force on the molecule and for the
continuum solvent all forces are Coulombic then
\[
\sum_{\alpha} {\rm Q}_{\alpha} {\bf E} (\boldr_{1\alpha}) = 0,
\]
the y-component of the field at the oxygen site and hydrogen sites
must be equal.
For the SPC simulations, there is an additional force on the oxygen atom
due to the non-bonded Lennard-Jones interaction which has a small nonzero
component in the y-direction (at Q$_H$=0.41, it is 1.47 kcal/mole/{\AA})
so the net force is zero but the electrostatic fields do not exactly balance.
The field from the md simulation
is not linear in charge, unlike the CT field.
This is consistent with the simulation
results for the electrostatic potential, $\phi$, being nonlinear and the
free energy, $\Delta {\rm A}_{es}$, being non-quadratic.
(Figures \ref{fig:phi}-\ref{fig:field}).
A non-quadratic charge dependence was
reported by Jayaram, et al\cite{Jayaram}, for the
solvation of a spherical cation.

Any conclusions regarding the validity of
dielectric continuum theory are of course
dependent on the input parameters, particularly on the radii, ${\rm r}_O$ and
${\rm r}_H$. The values for the $\Delta {\rm A}_{es}$, the x- and y-components
of the electric field and the Coulomb energy, $\langle {\rm U}_{es}
\rangle$,
are given in Table \ref{tab:compare}.
(From the fact that the electrostatic
potential $\langle \phi \rangle$ is linear in charge, it follows
from Eqs. \ref{eq:ues} and \ref{eq:dg} that $\Delta {\rm A}_{es}
= \langle {\rm U_{es}} \rangle$ /2 for the continuum results.)
Also shown on Table \ref{tab:compare} are the SPC simulation results
and the results of two approximate models (see below).
Some cavity radii
give accurate estimates for the free energy,
but it is not possible for continuum theory to simultaneously
predict values of $\Delta {\rm A}_{es}$, $\langle {\rm U}_{es} \rangle$,
and the electric field, {\bf E}, to all within 20$\%$.
In addition, the charge dependence of $\langle {\rm U}_{es} \rangle$
and {\bf E}
must be linear and $\Delta {\rm A}_{es}$
must be quadratic, in contrast to the results of the simulation.
There are sets of radius parameters which will give the best quadratic
fit to the simulation free energy. Two such sets are
(r$_O$=1.88 {\AA}, r$_H$=0.80 {\AA}) and
(r$_O$=1.83 {\AA}, r$_H$=1.00 {\AA}) which will give the free energy curve
shown by the dotted line of Figure \ref{fig:energy}.

The solvent reorganization that results from increasing the charge
of the solute molecule is reflected in the radial distribution of charge,
defined by
\begin{equation}
{\rm Q}_{\alpha}({\rm r}) = {4 \pi \over 3 } \rho {\rm r}^2
\sum_{\beta=1}^3 {\rm Q}_{\beta} {\rm g}_{\alpha \beta} ({\rm r})
\label {eq:cdf}
\end{equation}
where $\rho$ is the bulk density of the solvent, the factor of three in the
dominator is introduced to reflect that there are three atomic sites,
and ${\rm g}_{\alpha \beta}$
is the pair correlation function between the atomic sites $\alpha$ on the
solute and $\beta$ on the solvent molecules\cite{Levy}.
Figure \ref{fig:cdf} shows Q$_O$(r)---the radial distribution of charge
as a function of distance from the solute oxygen site---for
three values of the solute charge: Q$_H$=0,0.25, and 0.50.
The distribution functions oscillate between positive hydrogen atom
peaks and negative oxygen atom peaks.
As the charges on the solute are increased,
the first peak moves in.
In addition, a peak grows in at a O-H hydrogen bond distance of 1.8 {\AA}.
This hydrogen-bond peak is faintly visible at Q$_H$=0.25 and is
prominent at Q$_H$=0.50.

We focus on two of the possible explanations
of the breakdown of dielectric continuum models:
1) dielectric saturation, in which the orientational
ordering (from hydrogen bonds) of the first solvation shell decreases
the dielectric response of the solvent and 2) electrostriction,
in which the solvent molecules come into closer
contact with the solute molecule as its polarity is
increased, and which thus leads to a decrease in the solute cavity
as solute charge is increased.
Dielectric saturation effects decrease the solvation energy
whereas electrostrictive effects increase the solvation energy.
Both of these solvent responses are visible in Figure \ref{fig:cdf}.
We will take two approaches to understanding the differences
in the md simulation free energies and the continuum free energies.
In one, we will scale the cavity size with solute charge in the continuum
model.
In the other, we will explicitly include first shell
waters in the CT calculations.

\vspace{5 mm}
\noindent
{\bf Scaled Radius}.
{}From Figure \ref{fig:cdf}, for
Q$_H$=0 the cavity is larger than $2^{-5/6} \sigma$ (=1.78 {\AA}),
since the solvent peak does not start until after 2 {\AA},
and the CT results, which use 1.77 as a cavity radius,
therefore overestimate the
free energy at low charges (see Figures \ref{fig:energy}).
This raises a question as to whether the differences between the
simulation and CT calculations are due to the approximation of a
continuum solvent or just an inconsistent choice of cavity size.
Estimates of the cavity size can perhaps be found from
an examination of the liquid structure, although assigning a sharp
solute/solvent boundary from a continuous distribution is ambiguous.
A simple heuristic method is to find the radius which
gives the best value for the Coulomb energy, $\langle {\rm U}_{es}\rangle$.
For each value of Q$_H$ the optimal oxygen radius is found
(the hydrogen radius is set equal to a constant value of 0.8 {\AA})
and the free energy is calculated from Eq. \ref{eq:dg}.
The resulting $\langle {\rm U}_{es}\rangle$
and $\Delta {\rm A}_{es}$ are in almost
exact agreement with the simulation values (see Table \ref{tab:compare}).
Of course, this agreement is by construction and
with one adjustable parameter it is a trivial accomplishment to
fit the simulation energies.
However, this method also gives good values for the electric
fields, which is not by construction(see Figure \ref{fig:fr0}).
The optimized oxygen radius is shown for each value of the hydrogen charge
in Figure \ref{fig:rO} (there is no optimized value at Q$_H$=0 since at
this point, $\langle {\rm U}_{es}\rangle$=0).
The oxygen radius shows a strong charge
dependence---the radius varies by 30$\%$ over the range of charges---and
this dependence is approximately linear with a slope of about 1.2 {\AA}/e.
There is some uncertainty in the radius for low values of the charge
since the
energy is small and variations of $\pm$ .1 {\AA} in ${\rm r}_O$ give
rise to energies which are all within the error bars of the simulation.
The values of the scaled ${\rm r}_O$ are shown by the arrows on Figure
\ref{fig:cdf} indicating
that these radii are consistent with the liquid structure.
The arrow on the Q$_H$=0 plot is the radius for Q$_H$=0.05.

\vspace{5 mm}
\noindent
{\bf Semicontinuum Methods}.
In this model the nearest neighbor solvent water molecules are treated
explicitly and the rest are treated as a continuum.
This method could include both
dielectric saturation effects as the first solvation shell
orders around the solute and electrostriction effects as the first
solvation shell moves closer to the solute.
The focus of this analysis is to see if we can explain the
non-linear effects seen in the simulation
in terms of reorganization of the local solvation shell.
This method is implemented as follows.
Configurations generated
from the SPC simulation are taken and the coordinates of the
solute plus its $n$ nearest neighbors
are inputed into the
DelPhi program. This is done for about 300 configurations, each taken
.1 picoseconds apart.
The energy and fields on the central solute will now be a sum of a
part coming from the $n$ explicit nearest neighbors and the dielectric
continuum.
This model is similar to previous semicontinuum methods, although here
1) we are averaging over configurations of the first shell molecules and
{\it not} using thermodynamic data for the first shell contribution as is
done in the other studies, and 2)
we are taking an arbitrary surface for regions beyond the first shell
(as defined by the radii of the first shell atoms) and {\it not} a
spherical shell{\cite{GoldmanBates,Pitzer}}.
Another semicontinuum study by Rashin and Bukatin includes one first shell
water inside the continuum and thermodynamic properties are found by
integrating
numerically over the solute and single solute water coordinates and then
extrapolating to a full hydration shell\cite{RashinBukatin}.
We are treating the number of explicit water molecules,
$n$, as a variable, ranging from 4 to 8,
in order to measure how many solvent molecules are strongly influenced
by the solute,

Figure \ref{fig:eqsc} compares the results of the full simulation.
the continuum theory (with ${\rm r}_O$=1.77 {\AA} and ${\rm r}_H$=0.80 {\AA})
and the semicontinuum results with 4 and 8 neighbors.
This figure gives $\Delta {\rm A}_{es}$ divided by Q$_H$ versus
Q$_H$.
The semicontinuum results and the full simulation exhibit
a similar non-linear
dependence on $\Delta {\rm A}_{es}/{\rm Q}_H$ on Q$_H$,
although the deviations from linearity are not as great for the
semicontinuum model.
This means that the non-linear effect is largely local, as also supported
by the scaled radius analysis.
At low charges the {\it n}=4 semicontinuum results are essentially the same as
the pure continuum results, implying that we have not included enough explicit
neighbors.
At high charges the semicontinuum results are different from the
continuum results.
The free energy at a given charge depends on values of the potential
at lower charges, through Eq. (\ref{eq:dg}).
Continuum theory with radii r$_O$=1.77{\AA} and r$_H$=0.8{\AA}
overestimates $\langle {\rm U}_{es} \rangle$ at low charges (because the oxygen
radius is too
small, see the preceding section) and
underestimates it at higher charges (see Table \ref{tab:compare}).
The $n$=4 semicontinuum results also overestimate the potential
energy at low charges (because it has not included enough
neighbors and is similar the pure continuum results) and at
higher charges, it does not have the same fortuitous and compensating
underestimation of $\langle {\rm U}_{es} \rangle$
(because at higher charges four
solvating waters can more adequately describe the local environment).
Therefore, the continuum theory gives better estimates of the free energy
at high charge than the $n$=4 semicontinuum method.
For other properties, such as $\langle {\rm U}_{es}\rangle$ (Table
\ref{tab:compare}) and
the electric fields (Table \ref{tab:compare} and Figure \ref{fig:fsc}),
the semicontinuum methods are closer to the simulation results
and as $n$ increases, the agreement improves.
At Q$_H$=.41 the first solvation shell contains 5 neighbors as measured by
integrating the $g_{OO}$ out to the first minimum.
At Q$_H$=0, the first solvation shell is broader and contains more molecules
(about 16). By including up to 8 explicit neighbors in the semicontinuum model
we are then including all the first shell solvent molecules of the higher
charge solute, but only some of the first shell molecules of the lower charge
solute.
At the highest charges, above .41, the semicontinuum results are the same for
4 and 8 neighbors, so at these charges there is strong tetrahedral ordering
4 nearest neighbors are enough to describe the local structure.

\vspace{5 mm}
\noindent
{\bf Conclusion}

The simple example of solvation presented here is
that of a single water molecule with partial charges
on the atomic sites which are varied between zero and values greater than the
charges for water (a partial charge on the hydrogen atoms equal to 0.50e).
The continuum theory (CT)
calculation by definition shows a simple quadratic dependence on charge
for the free energy, $\Delta {\rm A}_{es}$,
and a linear dependence for
electrostatic potential, $\phi$ and
the electric fields, {\bf E}.
{}From the md simulations, the charge dependence of $\Delta {\rm A}_{es}$
is definitely non-quadratic and that of $\langle \phi \rangle$
and {\bf E} non-linear
(see Figures \ref{fig:energy}, \ref{fig:phi} and \ref{fig:field}).
The md simulations then indicate
that some type of solvent reorganization other than
the re-orientation of the electric moments of the solvent molecules is
occurring as the solute is charged. Because these effects
are important and because they are not included in the CT calculations,
continuum theory cannot simultaneously predict free energies, potential
energies and electric fields to within 20$\%$
no matter what radius parameters are used
(see Table \ref{tab:compare}).
There are sets of radius parameters which will predict the free energies
shown by the dotted line in Figure \ref{fig:energy}, which represents
the best quadratic fit to the simulation free energy.

Two simple models are suggestive of the types of solvent reorganizations
that are
important to this solvation process. These models also provide a method to
improve continuum theory.
In one model, the size of the solute cavity
is charge dependent (scaled radius)
and in the other model the system inside the dielectric continuum contains
the solute together with explicit neighboring solvent molecules
(semicontinuum).
The energies and free energies for all of the different methods
(md simulation, pure continuum theory, scaled radius,
and semicontinuum) are summarized in
Table \ref{tab:compare}.
As can be seen from this Table and also from
Figure \ref{fig:fr0}, the scaled
radius method provides a good estimate of the free energy
and the electric fields.
The agreement between the scaled radius and md simulation results for
the Coulomb energy, $\langle {\rm U}_{es} \rangle$,
is by construction since $\langle {\rm U}_{es} \rangle$ is used to
determine the scaled radius.
The success of the scaled radius calculations suggest that
electrostriction plays a more important role
than dielectric saturation of close-lying solvent molecules
in explaining the deviations
of the continuum model from the full molecular solvent.
The oxygen radius used in the scaled radius method
varies by a large amount (30$\%$) with the charges of the
solvent (see Figure \ref{fig:rO}).
This large decrease in the solute cavity is also seen in
the liquid correlation functions from the simulation (see Figure
\ref{fig:cdf}).

In conclusion, simple continuum theory, as usually implemented in
DelPhi calculations, is not capable of determining free energies
to better than 2 kcal/mole accuracy and gives inaccurate
predictions of electrostatic potentials and electric fields.
We believe that continuum theory can be improved to give better than
1 kcal/mole accuracy by adopting charge dependent radii to
allow for electrostrictive effects.
It remains to invent a
theoretical model which predicts how site radii should be
scaled with charge.

\vspace{5 mm}
\noindent
{\bf Acknowledgments}

This work was supported by a grant
from the National Institutes of Health (GM43340-01A1)
and was done at the NIH Biotechnology Resource Center at Columbia University.
We would like to thank Christian Cortis, Prof. Richard Friesner and Prof.
Barry Honig for useful discussions.

\vfill
\eject
\baselineskip=0.56\baselineskip

\bibliography{ref_delphimdcomp}
\clearpage
\begin{table}
\caption{Comparison of the md simulation results (first row) with
continuum theory (CT)
results for a variety of cavity radii (${\rm r}_O$, ${\rm r}_H$),
the scaled radius CT results and semicontinuum results with {\it n}
explicit neighbors.
The properties listed are the free energy, $\Delta {\rm A}_{es}$,
the Coulomb energy, $\langle {\rm U}_{es} \rangle$, both in kcal/mole
and the x-component, ${\rm E}_x$, and y-component, ${\rm E}_y$, of the
electric field in kcal/(mole e {\AA}) (see Figure 4) for a solute with
Q$_H$=0.41.
The numbers in parenthesis are two standard deviation error estimates.}
\vspace{5 mm}
\begin{tabular}{|l|llll|}\cline{1-5}
  & $\Delta {\rm A}_{es}$
& $\langle {\rm U}_{es} \rangle$ & ${\rm E}_x
({\rm r}_{H1})$ & ${\rm E}_y ({\rm r}_{H1})$ \\
\cline{1-5}
Simulation & -8.4(5) & -23.32(6) & 22.2(1) & 33.6(1) \\
CT (${\rm r}_O$=1.77,${\rm r}_H$=0.80) &-10.5 & -21.0 & 25. & 32. \\
CT (${\rm r}_O$=1.88,${\rm r}_H$=0.80) & -8.2 & -16.4 & 17. & 25. \\
CT (${\rm r}_O$=1.83,${\rm r}_H$=1.00) & -8.2 & -16.4 & 14. & 25. \\
CT (${\rm r}_O$=1.80,${\rm r}_H$=1.00) & -8.6 & -17.2 & 15. & 26. \\
CT (${\rm r}_O$=1.70,${\rm r}_H$=1.10) & -8.7 & -17.4 & 12. & 27. \\
CT (${\rm r}_O$=1.50,${\rm r}_H$=1.16) & -10.1 & -20.2 & 12. & 31. \\
Scaled radius CT & -8.4 & -23.6 & 26. & 35. \\
semicontinuum(n=4) & -11.3(3) & -26.1(1) & 26.1(7) & 39.2(4) \\
semicontinuum(n=6) & -10.3(4) & -25.2(4) & 24.0(7) & 37.6(5) \\
semicontinuum(n=8) &  -9.7(5) & -24.2(2) & 22.9(3) & 36.1(2) \\
\cline{1-5}
\end{tabular}
\label{tab:compare}
\end{table}

\clearpage

\noindent
Figure 1: Thermodynamic cycle for the solvation of water in water.

\vspace{5 mm}
\noindent
Figure 2: Free Energy, $\Delta {\rm A}_{es}$, from md simulations (solid line)
and DelPhi continuum calculations (dashed line) as a function of solute charge
(in kcal/mole). The dotted line is a quadratic fit to the simulation data.

\vspace{5 mm}
\noindent
Figure 3: The averags electrostatic Potential, $\langle \phi \rangle$,
in kcal/mole/e,
at the hydrogen (solid lines)
and oxygen (dashed lines) position, comparing the
SPC simulations (the lines showing data points and error bars) with
DelPhi continuum results (no point symbols).

\vspace{5 mm}
\noindent
Figure 4: Electric field, {\bf E}, in kcal/mole/e/{\AA}
in the x-direction
at the ${\rm H}_1$ hydrogen atom (top), and
in the y-direction at the hydrogen and oxygen atoms (bottom) comparing
md simulation (solid (H) and dotted lines(O))
and DelPhi continuum calculations (dashed lines). In the continuum
model E$_y$ at the H and O sites are equal, for
the simulation there is a small difference (see text).
See the upper left-hand corner
for the definition of the coordinate system.

\vspace{5 mm}
\noindent
Figure 5: Solvent charge distribution (determined from the md simulations)
about the solute oxygen site for a)
Q$_H$=0, b) Q$_H$=0.25 and c) Q$_H$=0.50. The arrows indicate the optimized
oxygen cavity radius.

\vspace{5 mm}
\noindent
Figure 6: As for Figure 4, but comparing md simulation (solid lines),
DelPhi continuum results with fixed radii (dashed lines), and scaled radii
continuum results (dotted lines).

\vspace{5 mm}
\noindent
Figure 7: Optimized oxygen cavity radius, ${\rm r}_O$, as a function
of solute charge.

\vspace{5 mm}
\noindent
Figure 8: Free energy divided by hydrogen charge, Q$_H$, for the full md
simulation
(long-dashed line), semicontinuum with 4 neighbors
(dotted line), semicontinuum with 8 neighbors (short-dashed line),
and the continuum theory (solid line).

\vspace{5 mm}
\noindent
Figure 9: As for Figure 4, but comparing simulation (solid lines) results with
semicontinuum results for 4 (dotted lines)
and 8 neighbors (dashed line).

\newpage
\begin{figure}[p]
\include{delphimdfig1}
\caption{}
\label{fig:cycle}
\end{figure}
\eject

\begin{figure}[p]
\include{delphimdfig2}
\caption{}
\label{fig:energy}
\end{figure}
\eject

\begin{figure}[p]
\include{delphimdfig3}
\caption{}
\label{fig:phi}
\end{figure}
\eject

\begin{figure}[p]
\include{delphimdfig4}
\caption{}
\label{fig:field}
\end{figure}
\eject

\begin{figure}[p]
\include{delphimdfig5}
\caption{}
\label{fig:cdf}
\end{figure}
\eject

\begin{figure}[p]
\include{delphimdfig6}
\caption{}
\label{fig:fr0}
\end{figure}
\eject

\begin{figure}[p]
\include{delphimdfig7}
\caption{}
\label{fig:rO}
\end{figure}
\eject

\begin{figure}[p]
\include{delphimdfig8}
\caption{}
\label{fig:eqsc}
\end{figure}
\eject

\begin{figure}[p]
\include{delphimdfig9}
\caption{}
\label{fig:fsc}
\end{figure}
\eject

\end{document}